\documentclass[a4paper]{VisionStyle}
\usepackage{epsfig}

\newcommand{\subj}{MSH14-6{\it 3}}
\newcommand{\xmm}{XMM-Newton}
\newcommand{\Te}{$kT_{\rm e}$}
\newcommand{\net}{$n_{\rm e}t$}

\newcommand{\NH}{$N_{\rm H}$}
\newcommand{\NHunit}{${\rm cm}^{-2}$}

\newcommand{\spex}{{\it SPEX}}

\newcommand{\mekal}{{\it mekal}}

\newcommand{\jetp}{JETP}
\newcommand{\apj}{ApJ}
\newcommand{\aap}{A\&A}
\newcommand{\nat}{Nat}

\def\citep#1{(\cite{#1})}
\def\citet#1{\cite*{#1}}
\begin{document}

\title{
Non-thermal bremsstrahlung as the dominant hard X-ray 
continuum mechanism for the supernova remnant
MSH14-6{\it3} (RCW 86)}

\author{Jacco Vink\inst{1,2},
Johan A. M. Bleeker, Jelle S. Kaastra, Kurt van der Heyden\inst{3},
Andrew Rasmussen\inst{1}, John Dickel\inst{4}
} 

\institute{
Columbia Astrophysics Laboratory, Columbia University, MC 5247, 550 W 120th street, New York, NY 10027, USA
\and
Chandra fellow
\and
SRON National Institute for Space Research, Sorbonnelaan 2, NL-3584 CA, 
Utrecht, The Netherlands
\and
Astronomy Department, University of Illinois, 1002 West Green Street, Urbana, IL 61801, USA
}

\maketitle 

\begin{abstract}
We present an analysis of the X-ray emission of the supernova remnant \subj, 
which was partially covered by four observations with \xmm.
The detection of Fe K emission at 6.4~keV, and the lack of
spatial correlation between hard X-ray and radio emission is evidence
against a dominant X-ray synchrotron component. We argue that
the hard X-ray continuum is best explained by 
non-thermal bremsstrahlung from a supra-thermal tail to an otherwise cool 
electron gas. 
The existence of low electron temperatures, required to explain the absence
of line emission,
is supported by low temperatures found in other parts of the remnant,
which are as low as $0.2$~keV in some regions.
\end{abstract}

\section{Introduction}
  
The lack of bright emission lines in the X-ray spectra of some shell-type
supernova 
remnants has, for the last few years, been attributed to the presence of 
X-ray synchrotron radiation.
The best known example, and the first shell-type remnant for which
X-ray synchrotron radiation was claimed, 
is the remnant of \object{SN 1006} \citep{jvink-B2:Koyama95}.

Indeed, the small equivalent widths of X-ray emission lines 
from the remnant \object{MSH14-6{\it 3}} (also named \object{RCW 86} or \object{G315.4-2.3}) 
based upon ASCA observations \citep{jvink-B2:Vink97}, 
was reinterpreted by \citet{jvink-B2:Borkowski01} as due to
the presence of a synchrotron component.
Additional support for this idea was provided by
an apparent spatial correlation between the hard, continuum dominated 
X-ray emission and the radio synchrotron emission.

However, the presence of Fe K line emission at 6.4~keV complicates the
interpretation of the hard X-ray continuum.
This emission is the result of
inner-shell ionizations of neutral or mildly ionized iron,
followed by K shell fluorescence. 
Under-ionization can also account for the lack of emission lines
from the helium and hydrogen like ionization stages of Ne, Mg, Si, S,
which
dominate the line spectra of young remnants like 
Tycho and Cas A.
These elements have a lower fluorescence yield than Fe.

Under-ionization can occur if the plasma is
far out of ionization equilibrium, or
it may be the result of a low electron temperature, possibly 
caused by insufficient electron-ion temperature equilibration \citep{jvink-B2:Itoh84}. 
In the latter case the presence of additional energetic
electrons is required for the Fe K emission,
which should also give rise to non-thermal bremsstrahlung.
Such non-Maxwellian particle distributions are thought to be the 
result of collisionless shock heating \citep{jvink-B2:Bykov99}

Here we present additional evidence for non-thermal bremsstrahlung based on
XMM-Newton observations of \subj. We refer to \citet{jvink-B2:Dyer02} for a 
different, Chandra, vision on the hard X-ray emission from this remnant.

\begin{figure*}[t]
\hbox{
	\epsfig{figure=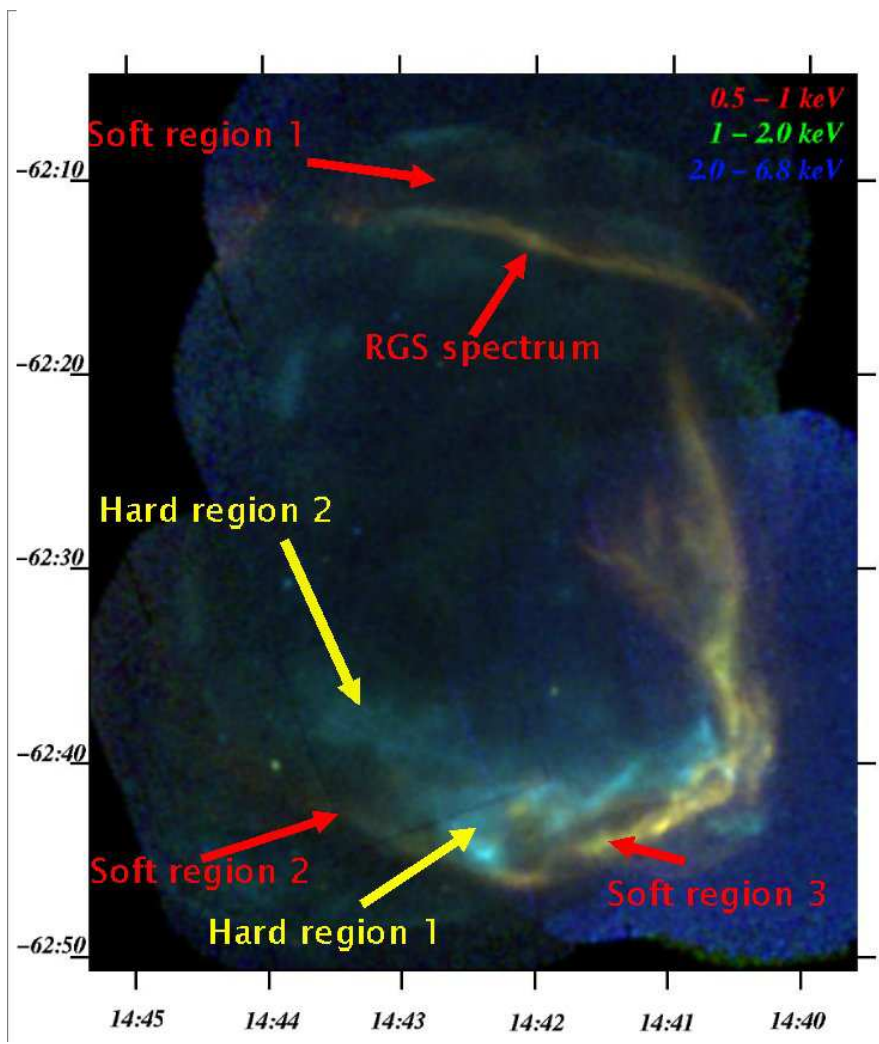,width=12.2cm}
	\hskip 1mm
\parbox[b]{55mm}{
	\epsfig{figure=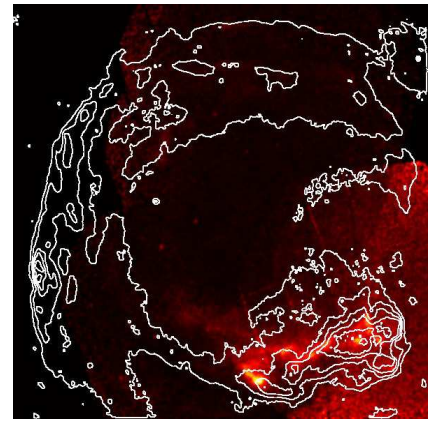,width=55mm}
	\vskip 15mm
\caption[]{Mosaic of exposure corrected EPIC PN and MOS images covering the 
SW, SE and NW of \subj. 
The color coding is as follows. Red : 0.54-1.0 keV; green: 1.0 -1.95 keV; 
blue: 1.95-6.8 keV. 
The labels are referred to in the text and in Fig.~\ref{jvink-B2_fig-spectra}. 
The figure on the right 
shows the 1.95-6.8 keV X-ray emission with
radio emission as observed by ATCA \citep{jvink-B2:Dickel00} overlayed in contours.
\label{jvink-B2_fig-mosaic}}
}}
\end{figure*}

\section{Observation and data reduction}
\xmm\  \citep{jvink-B2:xmm} observed the Southeast and Northwest of \subj\ 
on August 17 and 18, 2000, and the Southwest on August 18, 2001. 
The effective exposure times are approximately 12~ks for the Southwest,
17~ks for the Southeast, and 12~ks for the two
pointing towards the Northwest of the remnant.
We discuss here primarily the X-ray emission from the Southeast
and Northwest of the remant. The analysis of 
the Southwestern part of the remnant is still in a preliminary phase.
 
Most of the results presented here are based on data from the 
European Photon Imaging Camera (EPIC), consisting
of three CCD cameras MOS1 and MOS2,  and the PN. The latter
has the largest effective area, 
as it is behind the mirror that is not partially blocked by 
the reflective grating spectrometer (RGS), and the detector has a 
higher quantum efficiency \citep{jvink-B2:Strueder01}.
Observations were made with the medium-thick filter in place, which serves
to block ultra-violet contamination.

Satellite data were converted to cleaned event lists using SAS, 
the standard \xmm\ software package. 
Images and spectra were extracted from the event lists, using
only the standard event patterns 0 for PN, and 0-12 for MOS. This
allowed the use of standard spectral response matrices.

For the background correction of the spectra we used the publicly
available, standard, background event files. Unfortunately, these consist of
observations taken with the thin filter, possibly resulting in a background
overestimation at low photon energies. 
We therefore used additional off-center fields 
of the publicly available observation of G21.5-0.9, 
a small supernova remnant, which was observed with the medium filter. 
In practice we only used these data to estimate the background
for the soft X-ray emission from the Southeast of \subj.

\section{Data analysis}
\begin{figure*}[t]
\centerline{
\epsfig{figure=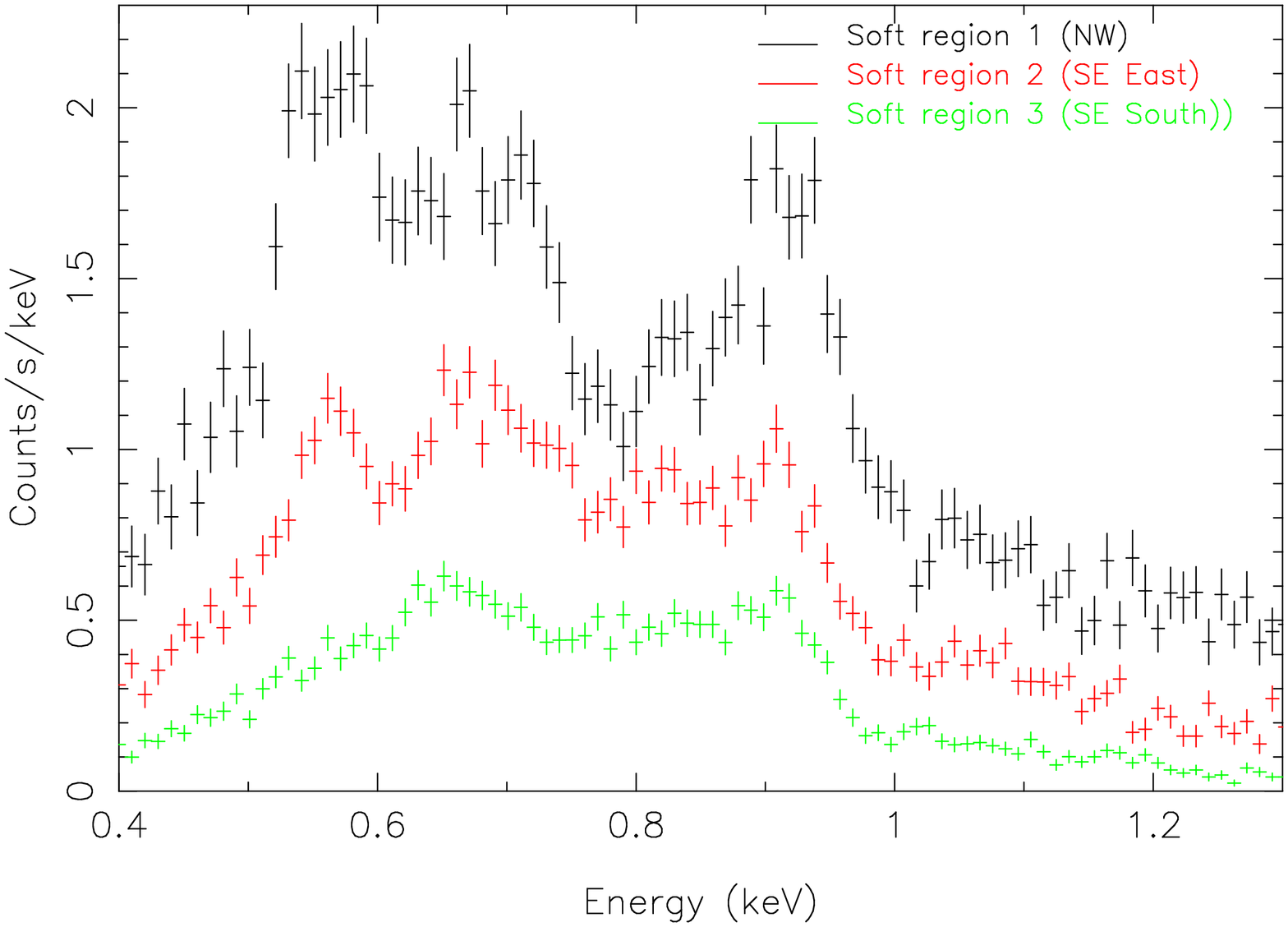,angle=-90,width=5.8cm}
\epsfig{figure=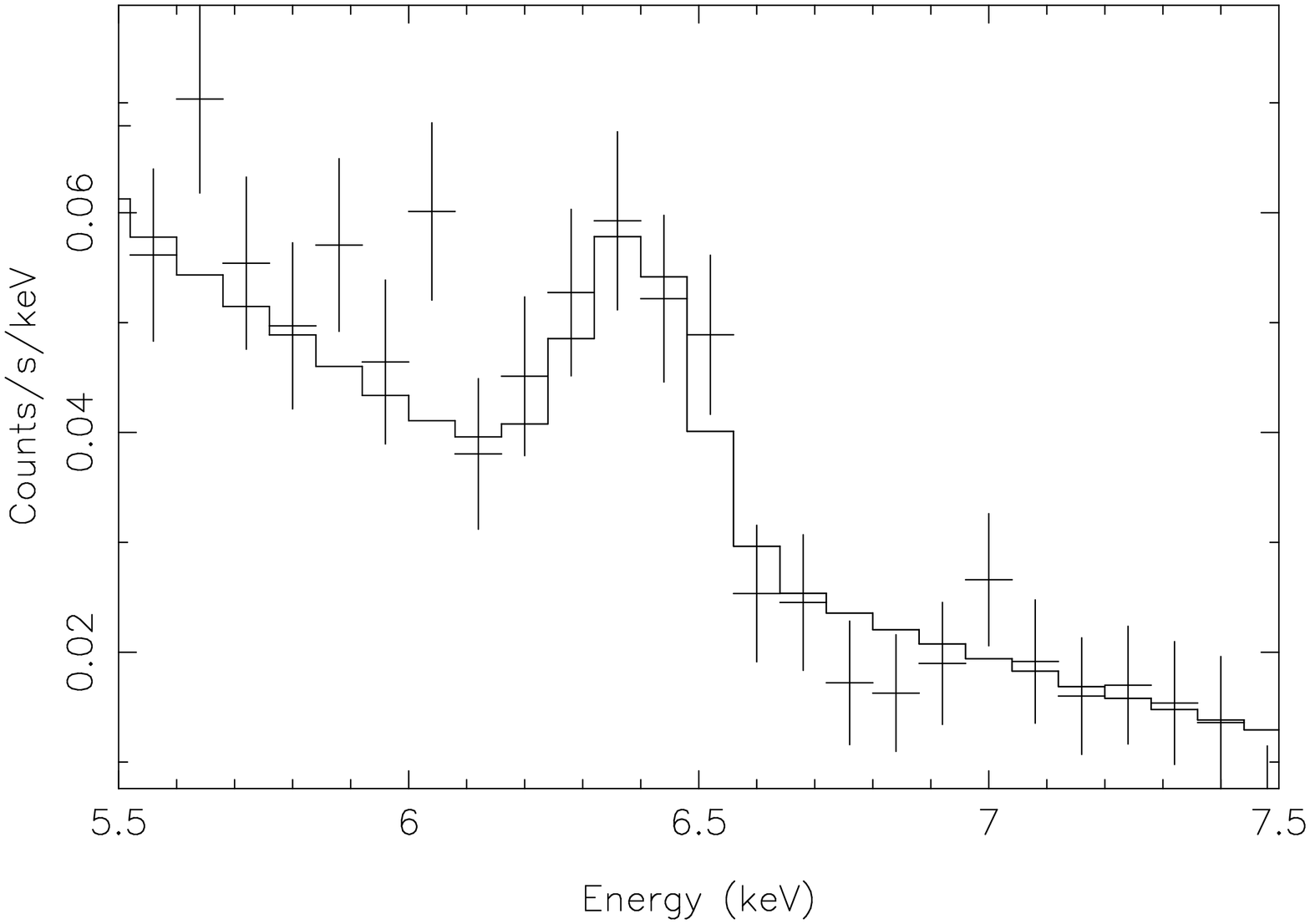,angle=-90,width=5.8cm}
\epsfig{figure=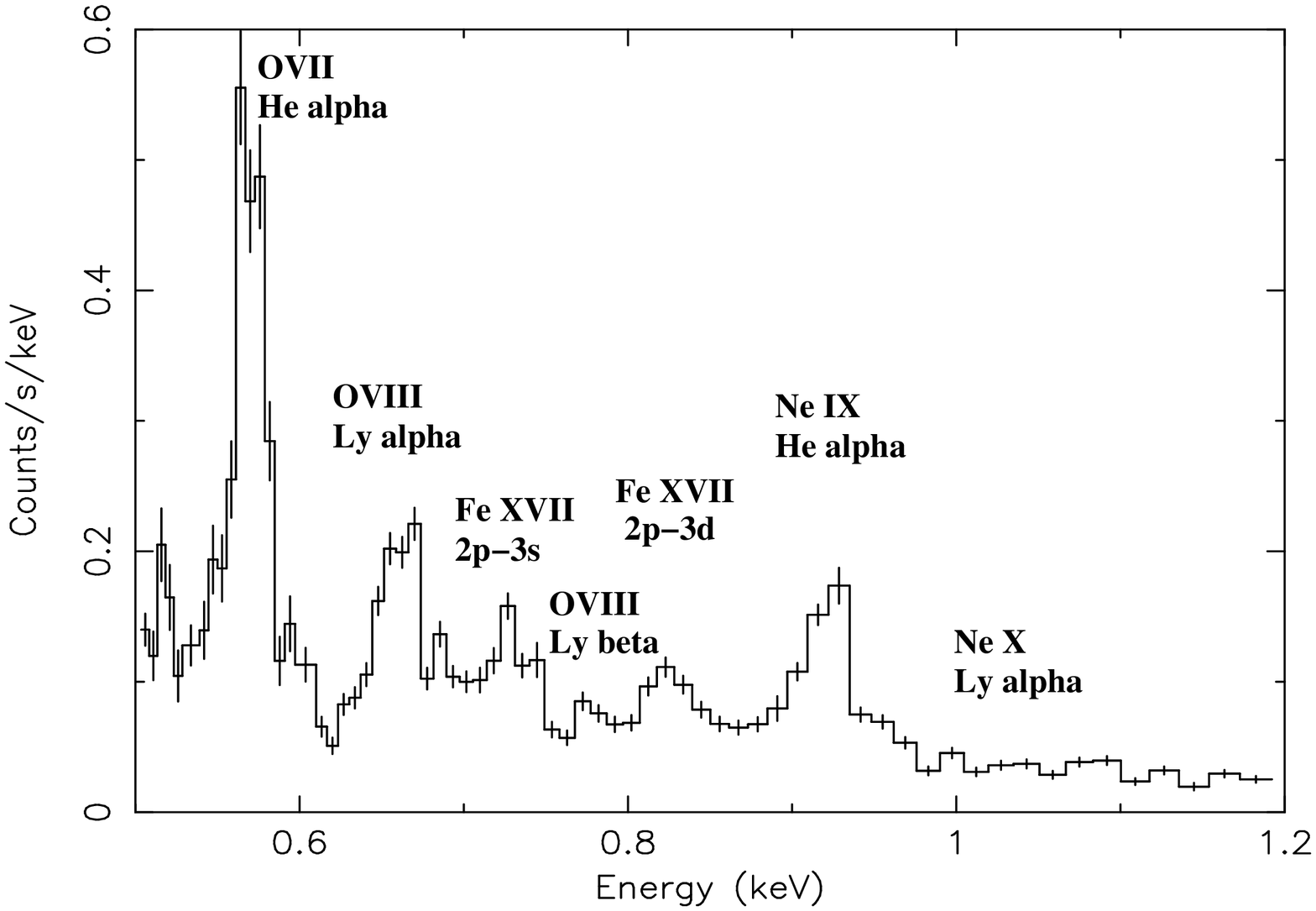,angle=-90,width=6.4cm}
}
\caption[]{
a) 
Three EPIC PN spectra arbitrary normalized to show a sequence of
increasing ionization from top to bottom. The labels refer to the
regions indicated in Fig.~\ref{jvink-B2_fig-mosaic}.
b) Part of the EPIC-PN spectra from the total hard X-ray emitting region in
the Southeast, showing Fe K emission around 6.4~keV, indicating
underionized iron.
c) The first order RGS spectrum of the Northwestern rim,
with labels identifying the line emission (cf. Fig.~\ref{jvink-B2_fig-spectra}a).
\label{jvink-B2_fig-spectra}}
\end{figure*}

Fig.~\ref{jvink-B2_fig-mosaic} shows a mosaic of the EPIC observations.
It illustrates the dramatic change in morphology 
going from energies of 0.5-1~keV to those $> 1$ keV, in more detail than 
previously observed by ASCA and BeppoSAX 
\citep{jvink-B2:Vink97,jvink-B2:Bocchino00}.
As shown in Fig.~\ref{jvink-B2_fig-spectra}a, the spectra of the soft X-ray regions
are dominated by emission lines of {O\,VII}, {O\,VIII}, {Fe\,XVII} 
and Ne IX, with some spatial variation in relative strength of the line 
emission. This is illustrated with better resolution by the RGS 
spectrum of the narrow, bright shell in the Northwest of \subj\
(Fig.~\ref{jvink-B2_fig-spectra}c). 

The presence of {O\,VII} is especially indicative of  a low ionization,
which is the result of a low electron temperature,
and/or a small ionization time\footnote{This is usually quantified by \net, 
where $n_{\rm e}$ is the electron density, 
and $t$ is the time since the plasma has been shocked.}.
A lower limit to the electron temperature is obtained by 
assuming collisional ionization equilibrium (CIE).
This gives temperatures for the
spectra in Fig.~\ref{jvink-B2_fig-spectra} between
0.10~keV and 0.17~keV, see Table~\ref{jvink-B2_tbl-soft}.\footnote{
The spectral range was confined to 0b.5-1 keV, which is dominated by
line emission. We used the \mekal\ CIE model \citep{jvink-B2:mekal},
and the \spex\ NEI model, which is based on the \mekal\ model.}
More realistic non-equilibrium ionization (NEI) models result in 
higher temperature estimates and somewhat better spectral fits 
(Table~\ref{jvink-B2_tbl-soft}). Although there are still discrepancies between
the NEI models and the data, 
possibly as a result of sharp density and temperature gradients and projection
effects, they indicate that at least
some regions have electron temperatures as low as 0.2~keV.

These temperatures are lower 
than indicated by the ASCA and the BeppoSAX 
data, but in accordance with the ROSAT PSPC data of 
the Northwestern rim \citep{jvink-B2:Bocchino00}. 
The electron temperatures are also consistent with shock 
velocities of 310 - 605~km/s, inferred from the width of the H$\alpha$\ and  
H$\beta$ lines, which, 
under the  assumption of full electron-ion equilibration 
\citep{jvink-B2:Ghavamian99,jvink-B2:Ghavamian01},
translates into shock velocities of 310 - 605~km/s,
and post shock gas temperatures of 0.11 - 0.42~keV.

Spectra from regions dominated by hard X-ray emission 
are characterized by a dominant 
continuum and very faint line emission.
The spectrum from ``hard region 1'', which is one of the brightest hard X-ray emitting regions, shows more prominent line emission than from 
``hard region 2''. It is not clear whether the more prominent line emission is
something intrinsic to this bright X-ray knot, or that this is caused by
a superposition of a soft X-ray emitting region, as this part of \subj\ has a
complicated morphology.
Good heuristic fits to the hard X-ray spectra consist of a power law 
continuum and a thermal components, but currently the spectral quality of this
very diffuse component is not good enough to determine for instance whether 
the emission lines come from a plasma out of ionization-equilibrium.
Clearly, the appearance of the hard spectra in itself can be explained
by X-ray synchrotron emission, with the line emission belonging to 
the thermal plasma, which is superimposed on the hard synchrotron continuum,
but the \xmm\ data show that there are two 
problems with this interpretation.

The first problem 
is that the hard X-ray emitting region in the Southeast as a whole
shows evidence for Fe K shell emission at 6.4~keV (Fig.~\ref{jvink-B2_fig-spectra}).
This confirms the ASCA detection of Fe K emission,
but for a different region, and with a lower equivalent width of 0.24~keV.
Model calculations show that for a solar iron abundance 
the measured equivalent width is consistent with electron temperatures
in excess of $\sim$3~keV, and/or a power law electron distribution with
an electron index $\la 3$.
\footnote{The calculations involved the numerical integration of an
electron distribution with the (almost) neutral iron cross sections, and
the bremsstrahlung gaunt factors proposed by \citet{jvink-B2:Haug97}.
The Lotz formula was used for Fe K shell ionization
cross sections,  and an approximate fluorescence yield of
0.3 was assumed (see e.g. \citet{jvink-B2:MeweEADN}).}
Therefore, only if the iron abundance is more than the solar value,
can the continuum be dominated by
X-ray synchrotron emission.

The second problem is the lack of detailed spatial correlation between hard X-ray and 
radio emission,
see for example (Fig.~\ref{jvink-B2_fig-mosaic} right)
the faint but distinct hard X-ray emission, labeled ``hard region 2'', 
which lies inside the radio shell.
This is in contradiction with the spatial correlation reported by
\citet{jvink-B2:Borkowski01}, based on lower spatial resolution data.
In fact, the region with the hardest X-ray emission
(hard region 2) has such weak radio emission, 
that it is difficult to determine the radio flux density from 
the ATCA and MOST images \citep{jvink-B2:Dickel00}, as these synthesis radio 
telescopes have sidelobes from brighter parts of the remnant which result 
in a negative flux density at this location.  
However, we have derived an upper limit of $< 0.5$ Jy at 1 GHz
based on these images and on a 5-GHz radio image made with the 
single-dish Parkes telescope \citep{jvink-B2:Milne75},
using the  generally accepted radio spectral index of -0.6 and total
flux density of 49~Jy at 1~GHz \citep{jvink-B2:Green-cat}.

Fitting the X-ray spectrum of this region with a power law 
spectrum and a thermal (CIE) spectrum yields a flux density
at 1 keV ($2.4\ 10^{17}$~Hz) of 1.6 x 10$^{-6}$~Jy.  
Connection of the radio and X-ray flux density 
gives a spectral index of $\ge -0.65$, which is remarkbly close
to the radio spectral index of -0.6, giving the fact that the two frequencies
encompass eight orders of magnitude. 
This does not rule out X-ray synchrotron emission,
but it is peculiar that for this particular region
synchrotron losses are almost absent,
whereas the radio regions are dominated by thermal soft X-ray emission; see for example
the northern shell, and the ridge of radio emission connecting
the southeastern and the northeastern shell, the region labeled
``soft region 2'' in Fig.~\ref{jvink-B2_fig-mosaic}.
Note that the hard X-ray regions lie inside the main
shell, which is a different spatial arrangement than for SN~1006,
where the X-ray synchrotron emission seems confined to a region close
to the shock front, and is highly limb brightened \citep{jvink-B2:Koyama95}.

\section{Summary and Conclusions}
We have presented spatially resolved \xmm\ spectroscopy of the supernova 
remnant \subj,
which shows that
the hard X-ray emission is not 
correlated with the radio emission. Fe K emission at 6.4~keV is also
present. Both observations contradict the idea that
the hard X-ray emission is dominated by synchrotron radiation.

The Fe K emission indicates the presence of electrons with 
energies in excess of the ionization threshold 
($\ga 7$~keV), but these electrons should also
give rise to bremsstrahlung.
A {\em non-thermal} electron distribution seems likely,
as a {\em thermal} distribution requires electron temperatures \Te$>$3~keV,
which is at odds with both the measured shock velocities \citep{jvink-B2:Ghavamian99}
and the electron temperatures of the regions dominated by
soft X-ray emission.

Not discussed so far is the lack of line emission from O VII,
which requires that the energetic electrons are part of
an otherwise very cool electron distribution (\Te $<$ 30 eV). 
Although we presented
evidence for very low temperatures this clearly deserves future
attention. Moreover, as most electrons have low energies it means that
the electron density is probably much higher than indicated by the X-ray data.
Low temperatures may be the result of insufficient
electron-ion temperature equilibration.
Evidence for weak electron-ion equilibration was obtained from
UV spectroscopy of SN 1006, indicating
$T_e \sim 0.05 T_i$, where $T_i$ is the ion temperature \citep{jvink-B2:Laming96}.
Recent calculations of ionization fractions for non-thermal electron 
distributions show that the ion fractions are quite similar to those of
the equivalent temperature of the Maxwellian part of the distribution,
provided that the plasma is not in ionization equilibrium, as is usually
the case in supernova remnants \citep{jvink-B2:Porquet01}.
Clearly, more model calculations are needed to assess the effects of
non-thermal electron distributions on the line emission.

The evidence for non-thermal bremsstrahlung, raises the
question whether other remnants may not have a similar X-ray component.
This may be true, but in general it is hard to judge from hard X-ray tails 
alone whether the emission is non-thermal bremsstrahlung or X-ray synchrotron 
emission. The situation in \subj\ is more fortunate, as the overall electron 
temperature is relatively cool, 
which causes the energetic electrons to betray themselves by causing
Fe K fluorescence emission.
Moreover, it is possible that we observe \subj\ during a special period in 
its evolution, as 
non-thermal electron distributions can only exist for a time scale 
comparable to the electron self-equilibration time scale 
($\tau_{\rm ee} \sim 10^{8}/n_{\rm e}$~s, see \cite{jvink-B2:Itoh84}). 
This suggests that the extreme 
properties of the X-ray emission from this remnant may be related to
a recent interaction of the blast wave with the steep density gradients
associated with the wall of a wind blown cavity, 
as suggested by \citet{jvink-B2:Vink97}.

Clearly, the nature of the X-ray emission and its relation to the 
environment of \subj\ is far from solved, as a comparison of this
study with \citet{jvink-B2:Dyer02} shows.
However, more detailed imaging of the Fe K emission and possible
similar inner shell ionization lines from Si K, may provide clear enough
signatures for either non-thermal bremsstrahlung or 
X-ray synchrotron emission.

\begin{table}
\caption[]{Summary of spectral fits to the spectra of Fig.~\ref{jvink-B2_fig-spectra}a.
The degrees of freedom were 47 for CIE models (i.e. no entry in column 3) and
46 for NEI models. Elemental abundances were assumed to be solar.
Labels refer to Fig.~\ref{jvink-B2_fig-mosaic}.
\label{jvink-B2_tbl-soft}}
\scriptsize
\begin{tabular}{llllll}\hline\noalign{\smallskip}
 & \Te\ &  $\log($\net$)$ & norm &
\NH & $\chi^2$ \\
&  keV  &  cgs & $10^{13}$ cm$^{-5}$ & $10^{21}$\NHunit\\
\noalign{\smallskip}
\hline
\noalign{\smallskip}
Soft 1 & $0.088^{+0.03}_{-0.01} $  & - & $114^{+136}_{-104} $ & $9.0 \pm 0.1$& 140\\
              & $ 0.14 \pm 0.01$& $11.6\pm 0.2$ & $3.3^{+1.5}_{-1.1}$  & $6.9 \pm 0.2$& 106\\
Soft 2 & $0.15 \pm 0.01$       & -  & $1.2 \pm 0.3$ & $6.7 \pm 0.2$ & 159\\
 & $ 0.37 \pm 0.10$ &   $10.3 \pm 0.2$    & $0.06^{+0.08}_{-0.03}$ & $5.1 \pm 0.5$& 60\\
Soft 3 & $ 0.17 \pm 0.1$        & - &  $1.6 \pm 0.6$ & $6.9 \pm 0.1$ & 105 \\
 & $ 1.1 \pm 0.6$   & $ 9.9\pm 0.1$   & $0.02^{+0.02}_{-0.01}$ & $4.2 \pm 0.5$ & 93  \\
\noalign{\smallskip}
\hline
\end{tabular}
\end{table}

\begin{acknowledgements}
JV is supported by the NASA 
through Chandra Postdoctoral Fellowship Award number PF0-10011
issued by the Chandra X-ray Observatory Center.
\end{acknowledgements}


\begin{thebibliography}{}
\bibitem[\protect\astroncite{Bocchino et al.}{2000}]{jvink-B2:Bocchino00}
	Bocchino, F., Vink, J., Favata, F., et al., 
	 2000, \aap, 360, 671
\bibitem[\protect\astroncite{Borkowski et al.}{2001}]{jvink-B2:Borkowski01} 
	Borkowski, K.J., Rho J., Reynolds S.P., Dyer K. 2001, \apj, 550, 334
\bibitem[\protect\astroncite{Bykov \& Uvarov}{1999}]{jvink-B2:Bykov99} 
	Bykov, A.M., Uvarov, Y.A., 1999, \jetp, 88, 465
\bibitem[\protect\astroncite{Dickel et al.}{2000}]{jvink-B2:Dickel00} 
	Dickel, J.R., Strom, R.G., 
	Milne, D.K. 2000, \apj, 546, 447
\bibitem[\protect\astroncite{Dyer et al.}{2002}]{jvink-B2:Dyer02} 
	Dyer, K., et al., 2002, these proceedings
\bibitem[\protect\astroncite{Ghavamian}{1999}]{jvink-B2:Ghavamian99} 
	Ghavamian, P. 1999, Ph.D. thesis, Rice University
\bibitem[\protect\astroncite{Ghavamian et al.}{2001}]{jvink-B2:Ghavamian01} 
	Ghavamian, P., Raymond, J., 
	Smith, R.C.,  Hartigan, P. 2001, \apj, 547, 995
\bibitem[\protect\astroncite{Green}{2000}]{jvink-B2:Green-cat}
	Green D.A. 2000, 
	`A Catalogue of Galactic Supernova Remnants (2000 August version)', 
	(http://www.mrao.cam.ac.uk/surveys/snrs/).
\bibitem[\protect\astroncite{Haug}{1997}]{jvink-B2:Haug97} 
	Haug, E. 1997, \aap, 326, 417
\bibitem[\protect\astroncite{Itoh}{1984}]{jvink-B2:Itoh84} 
	Itoh, H. 1984, ApJ, 285,601
\bibitem[\protect\astroncite{Jansen et al.}{2001}]{jvink-B2:xmm} 
	Jansen, F., et al. 2001, \aap, 365, L1
\bibitem[\protect\astroncite{Koyama et al.}{1995}]{jvink-B2:Koyama95} 
	Koyama, K., Petre, R., Gotthelf, E., et al.
	1995, \nat, 378, 255
\bibitem[\protect\astroncite{Laming et al.}{1996}]{jvink-B2:Laming96} 
	Laming, J. M., Raymond, J., McLauglin, B., Blair, W. 1996, 
	\apj, 472, 267
\bibitem[\protect\astroncite{Laming}{2001}]{Laming01b} 
	Laming, J. M. 2001, \apj, 563, 828
\bibitem[\protect\astroncite{Mewe et al.}{1995}]{jvink-B2:mekal}
	Mewe, R., Kaastra, J., 
        Liedahl, D.A. 1995,  Legacy 6, 16
\bibitem[\protect\astroncite{Mewe}{1999}]{jvink-B2:MeweEADN} Mewe, R. 1999, 
	in X-ray Spectroscopy in Astrophysics, 
	ed. J. van Paradijs and J. Bleeker (Springer-Verlag)
\bibitem[\protect\astroncite{Milne \& Dickel}{1975}]{jvink-B2:Milne75} 
	Milne, D. K., Dickel, J. R.
	1975, Aust.J.Phys., 28, 209
\bibitem[\protect\astroncite{Porquet et al.}{2001}]{jvink-B2:Porquet01} 
	Porquet, D., Arnaud, M., Decourchelle, A. 2001, \aap, 373, 1110
\bibitem[\protect\astroncite{Str\"uder et al.}{2001}]{jvink-B2:Strueder01} 
	Str\"uder, L., et al.
	2001, \aap, 365, L18
\bibitem[\protect\astroncite{Vink et al.}{1997}]{jvink-B2:Vink97} 
	Vink, J., Kaastra J., Bleeker, J. 1997, \aap, 328, 628
\end{thebibliography}
\end{document}